\documentclass[conference]{IEEEtran}
\IEEEoverridecommandlockouts
\usepackage{graphicx}
\usepackage{textcomp}
\usepackage{xcolor}
\def\BibTeX{{\rm B\kern-.05em{\sc i\kern-.025em b}\kern-.08em
    T\kern-.1667em\lower.7ex\hbox{E}\kern-.125emX}}
\usepackage{cite}
\usepackage{amsmath,amssymb,amsfonts,amsmath,psfig}
\usepackage{multirow}
\usepackage{subfig}
\usepackage{graphicx}
\usepackage{hyperref}
\usepackage{geometry}
\usepackage[algo2e]{algorithm2e}
\usepackage{algorithm}
\usepackage{algpseudocode}
\newcommand{\bi}{\begin{itemize}}
\newcommand{\ei}{\end{itemize}}

\newcommand{\be}{\begin{enumerate}}
\newcommand{\ee}{\end{enumerate}}
\newcommand{\bd}{\begin{description}}
\newcommand{\ed}{\end{description}}
\newcommand{\bc}{\begin{center}}
\newcommand{\ec}{\end{center}}
\newcommand{\bt}{\begin{tabbing}}
\newcommand{\et}{\end{tabbing}}
\newcommand{\bfig}{\begin{figure}}
\newcommand{\efig}{\end{figure}}
\newcommand{\beq}{\begin{equation}}
\newcommand{\beqarr}{\begin{eqnarray}}
\newcommand{\beqarrn}{\begin{eqnarray*}}
\newcommand{\eeq}{\end{equation}}
\newcommand{\eeqarr}{\end{eqnarray}}
\newcommand{\eeqarrn}{\end{eqnarray*}}
\newcommand{\bflr}{\begin{flushright}\vspace{-0.2in}}
\newcommand{\eflr}{\end{flushright}}
\newcommand{\bsub}{\begin{subequations}}
\newcommand{\esub}{\end{subequations}}
\newcommand{\barr}{\begin{array}}
\newcommand{\earr}{\end{array}}

\def\BibTeX{{\rm B\kern-.05em{\sc i\kern-.025em b}\kern-.08em
    T\kern-.1667em\lower.7ex\hbox{E}\kern-.125emX}}
\begin{document}
\title{Adversarial Attacks on Deep Learning Based Power Allocation in a Massive MIMO Network}
\author{\IEEEauthorblockN{B. R. Manoj$^{\dagger}$, Meysam Sadeghi, and Erik G. Larsson$^{\dagger}$}
\IEEEauthorblockA{$^{\dagger}$Department of Electrical Engineering (ISY), Link{\"o}ping University, Link{\"o}ping, Sweden\\
Emails: $\text{manoj.banugondi.rajashekara@liu.se}, \text{m.sadeghee@gmail.com}, \text{erik.g.larsson@liu.se}$}
\thanks{This work was supported in part by Security-Link.}
}
\maketitle
\begin{abstract}
Deep learning (DL) is becoming popular as a new tool for many applications in wireless communication systems. However, for many classification tasks (e.g., modulation classification) it has been shown that DL-based wireless systems are susceptible to adversarial examples; adversarial examples are well-crafted malicious inputs to the neural network (NN) with the objective to cause erroneous outputs. In this paper, we extend this to regression problems and show that adversarial attacks can break DL-based power allocation in the downlink of a massive multiple-input-multiple-output (maMIMO) network. Specifically, we extend the fast gradient sign method (FGSM), momentum iterative FGSM, and projected gradient descent adversarial attacks in the context of power allocation in a maMIMO system. We benchmark the performance of these attacks and show that with a small perturbation in the input of the NN, the white-box attacks can result in infeasible solutions up to 86\%. 
Furthermore, we investigate the performance of black-box attacks. 
All the evaluations conducted in this work are based on an open dataset and NN models, which are publicly available.
\end{abstract}
\begin{IEEEkeywords}
Adversarial attacks, deep learning, massive MIMO, neural networks, power allocation, wireless security.
\end{IEEEkeywords}
\IEEEpeerreviewmaketitle
\section{Introduction}
In recent advances, deep learning (DL) models have shown the potential in the learning ability being computationally efficient for the machine learning tasks such as in data science, computer vision, and natural language processing \cite{Schmidhuber_dl_2015}. The significant success of DL in these tasks has created an increasing interest to employ DL in wireless communication systems. 
Resource allocation \cite{Sanguinetti_dl_2018}, channel encoding and decoding \cite{Liang_an_iterative_2018}, sensing and localization \cite{Arnold_novel_massive_2019, Bast_CSI_based_2020}, radio signal classification 
\cite{Sadeghi_adversarial_attacks_2019,Kim_over_the_2020,Filipovic_adversarial_examples_2019,
Flowers_evaluating_adversarial_2020,Bair_on_the_2019}
and end-to-end communications \cite{Sadeghi_physical_adversarial_2019}, are just a few examples of different applications of DL-based wireless systems. Despite this surge of interest in DL-based wireless systems, it has been shown that these systems are exposed to new security issues \cite{Sadeghi_adversarial_attacks_2019,Kim_over_the_2020,Filipovic_adversarial_examples_2019,
Flowers_evaluating_adversarial_2020,Bair_on_the_2019,
Sadeghi_physical_adversarial_2019,
Hameed_the_best_2020,
Erpek_deep_learning_2019,
Sagduyu_adversarial_deep_2019,
Shi_over_the_2020, Davaslioglu_Trojan_attacks_2019}. One class of such security issues are adversarial attacks, where the perturbation is added to the input of the neural network (NN) and causes erroneous classification \cite{Sadeghi_adversarial_attacks_2019,Kim_over_the_2020,Filipovic_adversarial_examples_2019,
Flowers_evaluating_adversarial_2020,Bair_on_the_2019,Sadeghi_physical_adversarial_2019}. These perturbations are small enough which are well-crafted by the adversaries to fool the NN with high probability. 
Based on the type of threat, adversarial attacks can be classified into white-box and black-box attacks; in white-box attacks, the adversary is assumed to have the complete knowledge of trained NN model (includes model architecture and parameters), whereas in black-box attacks, the adversary is assumed to have no access or limited knowledge to the trained NN model \cite{Sadeghi_physical_adversarial_2019}.

The existing body of works on adversarial attacks in DL-based wireless systems have mainly focused on {\em{classification-based}} applications (e.g., modulation classification \cite{Sadeghi_adversarial_attacks_2019,Kim_over_the_2020,Filipovic_adversarial_examples_2019,
Flowers_evaluating_adversarial_2020,Bair_on_the_2019}), and the {\em{regression-based}} applications (e.g., power allocation \cite{Sanguinetti_dl_2018}) are overlooked. In this paper, we try to address this gap by studying the adversarial attacks against DL-based optimal power allocation in a massive multiple-input-multiple-output (maMIMO) system. To enable reproducibility of our results, we use the openly available dataset and the NNs of \cite{Sanguinetti_dl_2018}. 

The main contributions of this paper are as follows:
\begin{itemize} 
\item[(i)] We use the common adversarial attacks in computer vision and image classification (fast gradient sign method (FGSM) \cite{Sadeghi_adversarial_attacks_2019}, momentum iterative FGSM (MI-FGSM) \cite{Dong_boosting_adversarial_2018}, and projected gradient descent (PGD)  \cite{Madry_towards_deep_2018}), and extend them for a regression problem in the context of a maMIMO wireless system. More specifically, in this work, we have considered the optimal power allocation problem in the downlink of a maMIMO network using maximal-ratio (MR) and multicell-minimum mean square error (M-MMSE) precoding schemes.
\item[(ii)] We benchmark our proposed adversarial attacks and show that PGD is the most successful attack in fooling the NN.
We also study both white-box and black-box attacks.  For the white-box attack, we demonstrate that the adversary can cause up to 86\% infeasible solutions to be generated by the proposed models in \cite{Sanguinetti_dl_2018}, just by adding a small perturbation in the input of NNs. 
\item[(iii)] We further analyze the effect of transferability of attacks using different precoding schemes, i.e., the success rate of black-box attacks. 
\end{itemize}

The rest of the paper is organized as follows. In Section II, we briefly describe the dataset details and the DL models. Section III presents the overview of adversarial attacks including the threat model and the attacks such as FGSM, MI-FGSM, and PGD for the considered regression problem in a wireless system. Experiments and the investigation of white-box and black-box attacks are discussed in detail in Section IV and Section V concludes the paper. 
\section{Dataset and the Deep Learning Model}
To investigate the reliability and the robustness of a regression-based DL model in wireless networks, we choose to use the dataset that is made publicly available for DL-based power allocation in multi-cell maMIMO networks \cite{Sanguinetti_dl_2018}. In \cite{Sanguinetti_dl_2018}, the aim of using the DL model is to perform the optimal power allocation using max-product policy in the downlink of a maMIMO network. The network consists of $L$ cells, each cell having a base station (BS) of $M$ antennas and $K$ user equipments (UEs). The downlink signal transmitted by the BS in cell $j$ is given by 
${\bf{x}}_j = \sum_{k=1}^{K} {\bf{w}}_{jk} \, \varsigma_{jk}$, where $ \varsigma_{jk} \sim {\cal{N}}(0, \rho_{jk})$ is the data signal in the downlink for user $k$ in cell $j$, with a precoding vector ${\bf{w}}_{jk} \in \mathbb{C}^M$ that determines the transmission beamforming satisfying $||{\bf{w}}_{jk}||^2 = 1$, thus, $\rho_{jk}$ is the transmission power. The optimal precoding schemes that have been adopted by the authors  \cite{Sanguinetti_dl_2018} in generating the dataset are MR and M-MMSE combining. The downlink signal-to-interference-plus-noise ratio (SINR) for a UE $k$ in cell $j$ is given as
\begin{equation}
\gamma_{jk}^{\text{dl}} = \frac{\rho_{jk} \; a_{jk}}{\sum_{l=1}^{L} \sum_{i=1}^{K} \rho_{li} \; b_{lijk} + \sigma^2} \, , 
\label{SINR_eq}
\end{equation}
where $\sigma^2$ is the additive white Gaussian noise (AWGN) variance and the average channel gain \cite{Sanguinetti_dl_2018} is 
\begin{equation}
a_{jk} = |\mathbb{E}\{{\bf{w}}_{jk}^{\text{H}} \, {\bf{h}}_{jk}^j \}|^2
\label{avg_ch_eq}
\end{equation} 
with  ${\bf{h}}_{jk}^j$  denoting the channel between BS $j$ and UE $k$ in cell $j$, $\mathbb{E}\{\cdot\}$ is the expectation operator, and $(\cdot)^{\text{H}}$ denotes the hermitian operation. The average interference gain  \cite{Sanguinetti_dl_2018} is 
\begin{equation}
b_{lijk} = 
\begin{cases}
     \mathbb{E}\{|{\bf{w}}_{jk}^{\text{H}} \, {\bf{h}}_{jk}^l|^2\},& (l,i) \neq (j,k) \\
    \mathbb{E}\{|{\bf{w}}_{li}^{\text{H}} \, {\bf{h}}_{jk}^l|^2\} - |\mathbb{E}\{{\bf{w}}_{li}^{\text{H}} \, {\bf{h}}_{jk}^j\}|^2,              & (l,i) = (j,k)
\end{cases}
\label{avg_interf_eq}
\end{equation}
In this work, we have considered the max-product SINR optimal power allocation strategy which is formulated as 
\begin{eqnarray}\label{max-prod_eq}
&& \underset{\rho_{jk}: \, \forall j, k}{\max}  \,\, 				\prod_{j=1}^{L} \prod_{k=1}^{K} \gamma_{jk}^{\text{dl}}\,,    \\ &&
	\text{s.t.}  \, \, \sum_{k=1}^{K} \rho_{jk} \leq P_{\mathrm{max}}^{\text{dl}}\, , \, j\in [1,L]\,,  
\end{eqnarray}
where $P_{\mathrm{max}}^{\text{dl}}$ is the maximum transmit power in the downlink. 
 
We define the dataset as $\{{\bf{x}}(n),{\bf{y}}(n)\}_{n=1}^{N_T}$, where ${\bf{x}}(n)$ is an input data sample with the output ${\bf{y}}(n)$, and $N_T$ is the dataset size. We denote a regression-based deep neural network (DNN) model as  $f(.;{\boldsymbol{\theta}})$, with input to the model as ${\bf{x}}$ and the predicted output 
as $f({\bf{x}}) = {\bf{y}}$, where ${\boldsymbol{\theta}}$ is the set of parameters of the model $f$. The corresponding loss function of $f(\cdot)$ is denoted as ${\cal{L}}({\boldsymbol{\theta}},\bf{x},\bf{y})$. For a given cell $j$, the goal of a DNN model is to learn the mapping between the geographical positions of UEs ${\bf{x}}$ ${\in \mathbb{R}}^{2KL}$ and the optimal power allocation solution ${\boldsymbol{\rho}}_j = [\rho_{j1}, \hdots, \rho_{jK}]$ which is obtained by solving 
(\ref{max-prod_eq}) using the knowledge of (\ref{avg_ch_eq}) and (\ref{avg_interf_eq}) through the traditional optimization techniques. 
The DNN model that is employed in the reference \cite{Sanguinetti_dl_2018} is a feedforward NN with fully connected layers, consisting of an input layer of dimension as $2KL$, $N$ hidden layers, and output layer of dimension as $K+1$. The output layer dimension is $K+1$ instead of $K$, due to the fact that NN learns the power constraint $\sum_{k=1}^{K} \rho_{jk}$ to improve the estimation accuracy. The values of all the parameters that are considered for the downlink of a maMIMO network are as follows: $L=4$, $K=5$, $M = 100$, $P_{\mathrm{max}}^{\text{dl}} = 500$mW, cell coverage area is $1$km $\times$ $1$km with an area of $250$m $\times$ $250$m for each cell, $\sigma^2 = -94$dBm, and the bandwidth = $20$MHz.

The dataset size during NN training phase is $N_T$ samples with the input UEs positions and the output power allocations as ${\bf{x}}(n)$ and ${\boldsymbol{\rho}}_j(n)$ for $j = 1, \hdots, L$, respectively. 
Each sample of input and output of NN are having dimensions of $40$ and $6$, respectively.  
During the test phase of DNN, the dataset size is $N_\text{test}$ samples, 
with input denoted as ${\bf{x}}_\text{t}(m)$,
where $m = 1, \hdots, N_\text{test}$, different from the training samples and the predicted power allocations of the DNN $f({\bf{x}}_\text{t}(m))$ as 
${\boldsymbol{\hat{\rho}}}_j(m) = [\hat{\rho}_{j1}, \hdots, \hat{\rho}_{jK}]$.  
The NN architecture that is considered for the max-product SINR power allocation with MR and M-MMSE precoding schemes \cite{Sanguinetti_dl_2018} is as shown in Table \ref{tab:table1} having trainable parameters of $6981$, which we refer to as `Model 1'. To further improve the estimation accuracy of DNN model in terms of average MSE as compared to that of in Table \ref{tab:table1}, a more complex NN is considered in \cite{Sanguinetti_dl_2018} as shown in Table \ref{tab:table2}. This NN architecture is having trainable parameters of $202373$, which we refer to as `Model 2'.

\begin{table}[t!]
  \begin{center}
    \caption{Model 1--NN architecture with trainable parameters of $6981$.}
    \label{tab:table1} \scalebox{0.9}{
    \begin{tabular}{|c|c|c|c|}  
        & Size & Parameters & Activation function \\
      \hline
      Input & 40 & - & - \\  \hline
      Layer 1 (Dense) & 64 & 2624 & elu \\ \hline
      Layer 2 (Dense) & 32 & 2080 & elu \\ \hline
      Layer 3 (Dense) & 32 & 1056 & elu \\ \hline
      Layer 4 (Dense) & 32 & 1056 & elu \\ \hline
      Layer 5 (Dense) & 5  & 165  & elu \\ \hline
      Layer 6 (Dense) & 6  & 36   & linear \\ \hline
    \end{tabular} }
  \end{center}
\end{table}
\begin{table}[t!]
  \begin{center}
   \vspace{-0.25em}
   \caption{Model 2--NN architecture with trainable parameters of $202373$.}
    \label{tab:table2} \scalebox{0.9}{
    \begin{tabular}{|c|c|c|c|}  
        & Size & Parameters & Activation function \\
      \hline
      Input & 40 & - & - \\  \hline
      Layer 1 (Dense) & 512 & 20992  & elu \\ \hline
      Layer 2 (Dense) & 256 & 131328 & elu \\ \hline
      Layer 3 (Dense) & 128 & 32896  & elu \\ \hline
      Layer 4 (Dense) & 128 & 16512  & elu \\ \hline
      Layer 5 (Dense) & 5   & 645    & elu \\ \hline
      Layer 6 (Dense) & 6   & 36     & linear \\ \hline
    \end{tabular}}
  \end{center}
\end{table}
\section{Overview of Adversarial Attacks}
In this section, we investigate the vulnerability of regression-based DNN models assuming that the adversary has full knowledge of the NN structure 
to generate the adversarial examples in a digital white-box manner; further, we analyze the black-box attacks. 

Given a DNN model $f({\bf{x}})$: ${\bf{x}} \in {\cal{X}} \rightarrow {\bf{y}} \in {\cal{Y}}$ that predicts ${\bf{y}}$ as an output with ${\bf{x}}$ being the input, where 
${\cal{X}} \subset	{\mathbb{R}}^u$ and ${\cal{Y}} \subset	{\mathbb{R}}^v$ with $u$ and $v$ are the dimensions of the inputs and outputs, respectively. The objective of the adversary is to generate the adversarial examples ${\bf{x}}_\text{adv}$ in the neighborhood of 
${\bf{x}}$ for a specific distance constraint such that the output of the DNN is erroneous. The adversarial example is crafted by adding a small perturbation to the original input data without changing the output such that 
$f({\bf{x}}) = {\bf{y}}$ and $f({\bf{x}}_\text{adv}) = {\bf{y}}^\text{adv}$, given the constraint that $||\eta||_p \leq \epsilon$, where ${\bf{y}}$ and ${\bf{y}}^\text{adv}$ are the predicted outputs by $f(\cdot)$ for the inputs ${\bf{x}}$ and ${\bf{x}}_\text{adv}$, respectively, $\eta = {\bf{x}}_\text{adv} - {\bf{x}}$, $\epsilon$ is the perturbation magnitude, $||\eta||_p$ is the $L_p$-norm of the adversarial perturbation which is bounded by $\epsilon$. 
\subsection{Threat Model}
In this subsection, we envision the threat model for the considered DL-based resource allocation system. After obtaining the position of each user in the cell, all the users transmit their positions (referred to as clean input ${\bf{x}}_\text{t}$) to a processing unit (database), which is then fed as input to the NN to allocate power for the users. 
We assume that the hostile entity referred to as the adversarial attacker can perturb the input that is fed to the NN. For example, one way of doing this is to employ
spoofing techniques, where the attacker is fed by some global navigation satellite system (GNSS) receivers that are very close to the users' positions \cite{Psiaki_gnss_spoofing_2016}, can change the position values of the users fed to the network. The objective of the attacker is to compute the adversarial perturbation of UEs positions in the direction of the gradient to increase the loss function such that the DL-based power allocation system is broken. Therefore, the attacker perturbs ${\bf{x}}_\text{t}$ such that the UEs positions are changed by a small amount $\eta$, i.e., ${\bf{x}}_\text{t} + \eta$ becomes the input to the NN. This perturbation is very small which is in the order of centimeters (cms) as compared to the actual position values. This can be physically interpreted as changing the UEs positions from the current position to a new position 
by an amount of $d_{\epsilon}$ (known as distance perturbation) 
in the direction such that the loss function is maximized. The loss function to be maximized is considered as sum of the user powers so that the NN outputs infeasible power solution and the DL-based system is completely collapsed. In this paper, in order to explore the impact of  adversarial attack methods on DL-based regression setting, the study focuses from the attacker point of view. 

Different methods for generating adversarial examples focus on maximizing some loss function caused by an adversarial perturbation under the constraint of $\epsilon$. The loss function is chosen such that the adversarial example generating technique exists allowing the adversary to choose an attack method to fool the DNN model with high probability and less computational cost.  
For the considered problem of power allocation in a maMIMO system using DL-based approach, generating ${\bf{x}}_\text{adv}$ for a given 
DNN trained model $f(\cdot)$ with original clean input data ${\bf{x}}_\text{t}$ can be achieved by using the following optimization problem as 
\begin{eqnarray}
\label{adv_optmiz}
&& \hspace{-10pt}\underset{{\bf{x}}_{\text{adv}}}{\text{argmin}} \left\lVert {\bf{x}}_{\text{adv}}- {\bf{x}}_\text{t}\right\rVert_p  \\ &&
\hspace{-10pt} \text{s.t.}  
\label{adv_optmiz_const_1}
\hspace{10pt} f({\bf{x}}_\text{adv}) = {\boldsymbol{\hat{\rho}}}_{j}^\text{adv} \,,  \,\, \sum_{k=1}^{K} {\hat{\rho}}_{jk}^\text{adv} > P_{\mathrm{max}}^{\text{dl}}\,, \,\, {\bf{x}}_{\text{adv}} \in {\cal{X}}\,, \\ && 
\hspace{10pt} \text{and}\,\, \left\lVert {\bf{x}}_{\text{adv}}- {\bf{x}}_\text{t}\right\rVert_p \leq \epsilon\,, 
\end{eqnarray}
where ${\hat{\rho}}_{jk}^\text{adv}$ is the adversarial power for the UE $k$ in cell $j$ predicted by the model $f(\cdot)$ for an adversarial perturbed input. For the given UEs positions ${\bf{x}}_\text{t}$, $f({\bf{x}}_\text{t})$ predicts power 
${\boldsymbol{\hat{\rho}}}_j = [\hat{\rho}_{j1}, \hdots, \hat{\rho}_{jK}]$ for all the UEs in cell $j$. The sum of predicted powers by the model $f({\bf{x}}_\text{t})$ of all the UEs in each cell $j$ is constrained to be less than 
$P_{\mathrm{max}}^{\text{dl}}$, i.e.,  $\sum_{k=1}^{K} {\hat{\rho}}_{jk} \leq P_{\mathrm{max}}^{\text{dl}}$. This we refer to as a feasible solution provided by the DNN model. If a small perturbation $\eta$ is added to 
${\bf{x}}_\text{t}$, then the new positions of the UEs ${\bf{x}}_{\text{adv}}$ (i.e., UEs position is displaced in some direction) is also within the given cell area, so we expect the model $f({\bf{x}}_{\text{adv}})$ to predict the optimal power allocations in the same order as $f({\bf{x}}_\text{t})$ 
as long as the constraint $||\eta||_p \leq \epsilon$ is satisfied. However, this does not hold, with adversarial perturbation, the model  
$f({\bf{x}}_{\text{adv}})$  predicts adversarial power 
${\boldsymbol{\hat{\rho}}}_j^\text{adv} = [\hat{\rho}_{j1}^\text{adv}, \hdots, \hat{\rho}_{jK}^\text{adv}]$, which can provide infeasible solutions, i.e.,  $\sum_{k=1}^{K} {\hat{\rho}}_{jk}^\text{adv} > P_{\mathrm{max}}^{\text{dl}}$, as shown in (\ref{adv_optmiz_const_1}) if we could maximize some loss function such that the system breakdowns. The adversarial examples that can be generated by using the optimization problem in (\ref{adv_optmiz}) is difficult to solve in practice as it is intractable. Thus, in order to craft adversarial examples efficiently in-terms of time consumption and computational cost, there are many sub-optimal methods that have been proposed in the literature \cite{Sadeghi_adversarial_attacks_2019}, \cite{Yuan_adversarial_examples_2019, Dong_boosting_adversarial_2018}. 

In this paper, we consider the class of gradient methods \cite{Yuan_adversarial_examples_2019, Dong_boosting_adversarial_2018, Madry_towards_deep_2018} with $L_{\infty}$-norm, i.e., 
adversarial perturbation constraint is given as $||\eta||_{\infty} \leq \epsilon$, which implies that $L_{\infty}$-norm is bounded  by perturbations with a radius of $\epsilon$. The reason behind choosing 
$L_{\infty}$-norm is that, since UEs position is the input to the NN, the adversarial perturbation to this input provides the maximum distance perturbation from the original UE position. The geometrical interpretation of this is that with $L_{\infty}$-norm constraint, the  perturbation corresponds to the maximum displacement of the UEs positions in the direction of gradient of loss function so that loss is increased (in the steepest direction). For an $L_{\infty}$-norm bound, the distance perturbation can be related to the magnitude of perturbation as $d_{\epsilon} = \sqrt{2}\, \epsilon$. 
\subsection{Fast gradient sign method (FGSM)}
\begin{algorithm}[!t]
	\label{alg_fgsm}
	\SetAlgoLined
	\textbf{Input:} $f(.;{\boldsymbol{\theta}})$,  ${\bf{x}}_\text{t}(m)$, $\epsilon = d_{\epsilon}/\sqrt 2$, $L$, $P_{\mathrm{max}}^{\text{dl}}$, $N_\text{test}$ \\
 	\textbf{Output:} ${\bf{x}}_j^\text{adv}(m)$, ${\boldsymbol{\hat{\rho}}}_{j}^\text{adv}(m) = [{\hat{\rho}}_{j1}^\text{adv}, \hdots, {\hat{\rho}}_{jK}^\text{adv}]$
	
	\For {$j$ $\mathrm{in}$ range($L$)}
	{
	 \For {$m$ $\mathrm{in}$ range($N_{\mathrm{test}}$)}
	 {
	 ${\boldsymbol{\hat{\rho}}}_j(m) = [\hat{\rho}_{j1},\hdots,\hat{\rho}_{jK}] \leftarrow f({\bf{x}}_\text{t}(m))$\\	 
	 ${\cal{L}}({\boldsymbol{\theta}}, f({\bf{x}}_\text{t}(m))) = \sum_{k=1}^{K} {\hat{\rho}}_{jk}$ \\
     $\eta = \epsilon \, {\text{sign}}\left(\nabla_{{\bf{x}}_\text{t}(m)} {\cal{L}}({\boldsymbol{\theta}}, f({\bf{x}}_\text{t}(m))) \right)$\\
     ${\bf{x}}_j^\text{adv}(m) = {\bf{x}}_\text{t}(m) + \eta$\\
     ${\boldsymbol{\hat{\rho}}}_{j}^\text{adv}(m) = [{\hat{\rho}}_{j1}^\text{adv}, \hdots, {\hat{\rho}}_{jK}^\text{adv}] \leftarrow f({\bf{x}}_j^\text{adv}(m))$\\
\eIf{$\sum_{k=1}^{K} {\hat{\rho}}_{jk}^\mathrm{adv} >  P_{\mathrm{max}}^{\mathrm{dl}}$}
        {
			${\boldsymbol{\hat{\rho}}}_{j}^\text{adv}(m) \leftarrow$ infeasible power
		}
		{
		${\boldsymbol{\hat{\rho}}}_{j}^\text{adv}(m) \leftarrow$ feasible power	 
		}
	 }
	 ${\bf{x}}_j^\text{adv} = [{\bf{x}}_j^\text{adv}(1), \hdots, {\bf{x}}_j^\text{adv}(m)]$\\
	 ${\boldsymbol{\hat{\rho}}}_{j}^\text{adv} = [{\boldsymbol{\hat{\rho}}}_{j}^\text{adv}(1), \hdots, {\boldsymbol{\hat{\rho}}}_{j}^\text{adv}(m)]$ 
	}
 \caption{Generating adversarial examples using FGSM}
\end{algorithm}
The FGSM is a one-step gradient-based attack approach that can generate computationally efficient adversarial examples. In FGSM, adversarial example ${\bf{x}}_\text{adv}$ is crafted by maximizing the loss function ${\cal{L}}({\boldsymbol{\theta}},f({\bf{x}}_\text{t}))$ of the DNN such that $L_{\infty}$-norm is bounded. The adversarial perturbation \cite{Sadeghi_adversarial_attacks_2019} can be expressed as 
\begin{eqnarray}
\eta = \epsilon \, {\text{sign}}\left(\nabla_{{\bf{x}}_\text{t}} {\cal{L}}({\boldsymbol{\theta}},f({\bf{x}}_\text{t})) \right) \, ,
\label{fgsm_perturb}
\end{eqnarray}
where $\nabla_{{\bf{x}}_\text{t}} {\cal{L}}({\boldsymbol{\theta}},f({\bf{x}}_\text{t}))$ is the gradient of ${\cal{L}}({\boldsymbol{\theta}},f({\bf{x}}_\text{t}))$ w.r.t. the clean input data ${\bf{x}}_{\text{t}}$. The adversarial example is generated as ${\bf{x}}_\text{adv} = {\bf{x}}_{\text{t}} + \eta$. 
Algorithm~$1$ presents our proposed cell-based FGSM attack against power allocation in a maMIMO network.
We have considered a cell-based attack, where  each cell is attacked by an adversary by maximizing the loss function of the respective cell in the given network. In Algorithm~$1$, ${\bf{x}}_\text{t} = [{\bf{x}}_\text{t}(1), \hdots, {\bf{x}}_\text{t}(m)]$, ${\bf{x}}_j^\text{adv} = [{\bf{x}}_j^\text{adv}(1), \hdots, {\bf{x}}_j^\text{adv}(m)]$, and ${\boldsymbol{\hat{\rho}}}_{j}^\text{adv}(m) = [{\hat{\rho}}_{j1}^\text{adv}, \hdots, {\hat{\rho}}_{jK}^\text{adv}]$ are the clean input samples, the adversarial perturbed samples in cell $j$, and the adversarial power allocation obtained as output of the DNN in cell $j$ for the sample $m$, respectively, where $m \in [1, N_\text{test}]$ and $j \in [1, L]$.
\subsection{Projected gradient descent (PGD) method}
The FGSM can be extended by running a much finer optimization in an iterative manner. The multi-step iterative variant of FGSM called the PGD attack is a more powerful attack \cite{Madry_towards_deep_2018}. The PGD performs FGSM with a small step size $\alpha$ and projects the updated adversarial sample learned from each iteration onto the $L_{\infty}$-ball around the original clean input sample (i.e., $\epsilon$-neighborhood of ${\bf{x}}_{\text{t}}$). The update rule for a $Q$-step PGD attack is defined in the $q$-th iteration as follows:
\begin{eqnarray}
&& \hspace{-3em} {\bf{x}}_{{\text{t}},0} = {\bf{x}}_{\text{t}}  \\
&& \hspace{-3em} {\bf{x}}_{\text{t},q+1} = {\text{clip}}_{[{\bf{x}}_{\text{t}}, \epsilon]} \{{\bf{x}}_{\text{t},q} + \alpha \, \text{sign} \left(\nabla_{{\bf{x}}_{\text{t},q}} {\cal{L}}({\boldsymbol{\theta}},f({\bf{x}}_{\text{t},q}))\right) \} \\ 
&& \qquad \qquad {\bf{x}}_{\text{adv}} = {\bf{x}}_{\text{t},Q} \,, 
\label{pgd_update}
\end{eqnarray}
where $\alpha$ is the small step size, $Q$ is the number of iterations, ${\text{clip}}_{[{\bf{x}}_{\text{t}}, \epsilon]}\{{\bf{x}}_{{\text{t}},q}\}$ is the element-wise clipping of ${\bf{x}}_{{\text{t}}, q}$ to $[{\bf{x}}_{\text{t}} - \epsilon, {\bf{x}}_{\text{t}} + \epsilon]$ so that the result will be in $L_{\infty}$ $\epsilon$-neighborhood of ${\bf{x}}$. The values of $\alpha$  and $Q$ are chosen in such a way that the adversarial sample reaches the edge of the $L_{\infty}$ $\epsilon$-norm ball while restricted to keep minimal computation cost. Algorithm~$2$, presents our proposed cell-based PGD attack against power allocation in a maMIMO network.
\begin{algorithm}[!t]
	\label{alg_pgd}
	\SetAlgoLined
	\textbf{Input:} $f(.;{\boldsymbol{\theta}})$,  ${\bf{x}}_\text{t}(m)$, $\epsilon = d_{\epsilon}/\sqrt 2$, $L$, $P_{\mathrm{max}}^{\text{dl}}$, $\alpha$, $Q$, $N_\text{test}$ \\
 	\textbf{Output:} ${\bf{x}}_j^\text{adv}(m)$, ${\boldsymbol{\hat{\rho}}}_{j}^\text{adv}(m) = [{\hat{\rho}}_{j1}^\text{adv}, \hdots, {\hat{\rho}}_{jK}^\text{adv}]$
	
	\For {$j$ $\mathrm{in}$ range($L$)}
	{
	 \For {$m$ $\mathrm{in}$ range($N_{\mathrm{test}}$)}
	 {		 
		 \For {$q$ $\mathrm{in}$ range($Q$)}
	 		{
	 		${\boldsymbol{\hat{\rho}}}_{j,q}(m) = [\hat{\rho}_{j1},\hdots,\hat{\rho}_{jK}] \leftarrow f({\bf{x}}_{\text{t},q}(m))$\\
		    ${\cal{L}}({\boldsymbol{\theta}}, f({\bf{x}}_{\text{t},q}(m))) = \sum_{k=1}^{K} {\hat{\rho}}_{jk}$ \\ 
		    $\eta_q =  {\text{sign}}\left(\nabla_{{\bf{x}}_{\text{t},q}(m)} {\cal{L}}({\boldsymbol{\theta}}, f({\bf{x}}_{\text{t},q}(m))) \right)$\\
		    ${\bf{x}}_{\text{t},q+1}(m) = {\bf{x}}_{\text{t},q}(m) + \alpha \, \eta_q$\\
		    ${\bf{x}}_{\text{t},q+1}(m) \leftarrow {\text{clip}}_{[{\bf{x}}_\text{t}(m), \epsilon]} \{{\bf{x}}_{\text{t},q+1}(m)\} $
	 
	 		}
	 		$ {\bf{x}}_j^\text{adv}(m) \leftarrow {\bf{x}}_{\text{t},Q}(m)$\\
    		${\boldsymbol{\hat{\rho}}}_{j}^\text{adv}(m) = [{\hat{\rho}}_{j1}^\text{adv}, \hdots, {\hat{\rho}}_{jK}^\text{adv}] \leftarrow f({\bf{x}}_j^\text{adv}(m))$\\
\eIf{$\sum_{k=1}^{K} {\hat{\rho}}_{jk}^\mathrm{adv} >  P_\mathrm{max}^{\mathrm{dl}}$}
        {
			${\boldsymbol{\hat{\rho}}}_{j}^\text{adv}(m) \leftarrow$ infeasible power
		}
		{
		${\boldsymbol{\hat{\rho}}}_{j}^\text{adv}(m) \leftarrow$ feasible power	 
		}
	 }
	 ${\bf{x}}_j^\text{adv} = [{\bf{x}}_j^\text{adv}(1), \hdots, {\bf{x}}_j^\text{adv}(m)]$\\
	 ${\boldsymbol{\hat{\rho}}}_{j}^\text{adv} = [{\boldsymbol{\hat{\rho}}}_{j}^\text{adv}(1), \hdots, {\boldsymbol{\hat{\rho}}}_{j}^\text{adv}(m)]$ 
	}
 \caption{Generating adversarial examples using PGD}
\end{algorithm}
\subsection{Momentum iterative fast gradient sign method (MI-FGSM)}
Dong et al. \cite{Dong_boosting_adversarial_2018} proposed a technique called MI-FGSM which incorporates momentum memory into FGSM across multiple iterations. The gradients in the $(i+1)$-th iteration are calculated as 
\begin{equation}
g_{i+1} = \mu \, g_i + \frac{\nabla_{{\bf{x}}_{\text{t},i}}{\cal{L}}({\boldsymbol{\theta}},f({\bf{x}}_{\text{t}, i}))}{||\nabla_{{\bf{x}}_{\text{t},i}}{\cal{L}}({\boldsymbol{\theta}},f({\bf{x}}_{\text{t}, i}))||_1}\,,
\label{grad_mifgsm_update}
\end{equation}
where $\mu$ is the decay factor and $||\cdot||_1$ is the $L_1$-norm.
The regression-based MI-FGSM algorithm is similar to the proposed Algorithm~$2$, except that the adversarial examples are updated as 
${\bf{x}}_{\text{t},i+1} = {\bf{x}}_{\text{t},i} + \beta \, {\text{sign}}(g_{i+1})$, until the $I$-th iteration and then the clipping operation after each iteration is not performed, where $\beta$ is the step size and $I$ is the number of iterations. The parameters are initialized as $g_0 = 0$ and $\beta = \epsilon/I$. 
\subsection{Transferability}
Transferability \cite{Papernot_transferability_in_2016} is a property where the adversarial examples that are generated against a NN model can fool other NN models with different architecture. The attackers can train a surrogate NN model which is then used to generate the adversarial examples against this surrogate model. Transferability is an important property for black-box attacks where the attacker does not have access to either the target DNN model or the training dataset. Because of transferability, the target DNN will be vulnerable to the adversarial examples generated using the surrogate model. In Section {IV}, we discuss the transferability in detail by generating the adversarial examples using the substitute `Model 1' and attack the DNN `Model 2' using $L_{\infty}$-norm adversary attacks. 
\section{Experiments and Discussions}
In this section, we analyze the vulnerability of the DNN `Model 1' and `Model 2' for the max-product SINR power allocation strategy in the downlink of a maMIMO network with MR and M-MMSE precoding schemes using the $L_\infty$-norm  of FGSM, MI-FGSM, and PGD adversary attacks. 
We consider the same dataset that has been used in\cite{Sanguinetti_dl_2018} and is publicly available in the provided link: \url{https://data.ieeemlc.org/Ds2Detail}. 
In the experiments, we consider $N_T = 329000$ and $N_\text{test} = 500$ samples. 
For the PGD attack, we have $\alpha = 0.01$ and $Q=40$. For the MI-FGSM attack, we have $\mu = 0.1$, $I = 10$, and $\beta = \epsilon/I = 0.1\,\epsilon$. 
The values of the remaining parameters such as $K$, $L$, $M$, $P_\text{max}^\text{dl}$, $\sigma^2$, and coverage area of each cell are defined in Section II. 
\subsection{Digital white-box attacks}
By implementing the adversarial attacks that are proposed in subsection III-B to III-D, we investigate and compare the potential of different attacks in terms of their success rate to fool the NN models. The experiments are conducted on $500$ adversarial samples as input to the DNN models and 
analyze the trained network that for how many of these adversarial samples, the network provides infeasible power solution as output. 
In the test samples $N_{\text{test}}=500$ that are considered in our experiment, without adversarial attack, we have tested for their feasibility and made sure that for all these test samples, the DNN  `Model 1' and `Model 2' provide only feasible output solution across all the cells, i.e., $\sum_{k=1}^{K} {\hat{\rho}}_{jk} \leq P_{\mathrm{max}}^{\text{dl}}$ which implies that for each clean input sample to the NN, the output of it with sum of the user's powers in each cell is less than $P_{\mathrm{max}}^{\text{dl}}$. 
Furthermore, we consider the perturbation of clean input samples with random white Gaussian noise. We define ${\bf{x}}_\text{rnd} = {\bf{x}}_\text{t} + \epsilon \, \text{sign}({\bf{w}})$, where ${\bf{w}} \sim {\cal{N}}\left(0,\mathbf{I}\right)$, as 
the perturbation of UEs positions in the random directions with the same perturbation magnitude $\epsilon$ as that is considered while generating the adversarial examples. 
 
\begin{figure}[!htp]
\begin{minipage}[b]{0.493\linewidth}
  \centering
  \centerline{\includegraphics[width=1.13\linewidth]{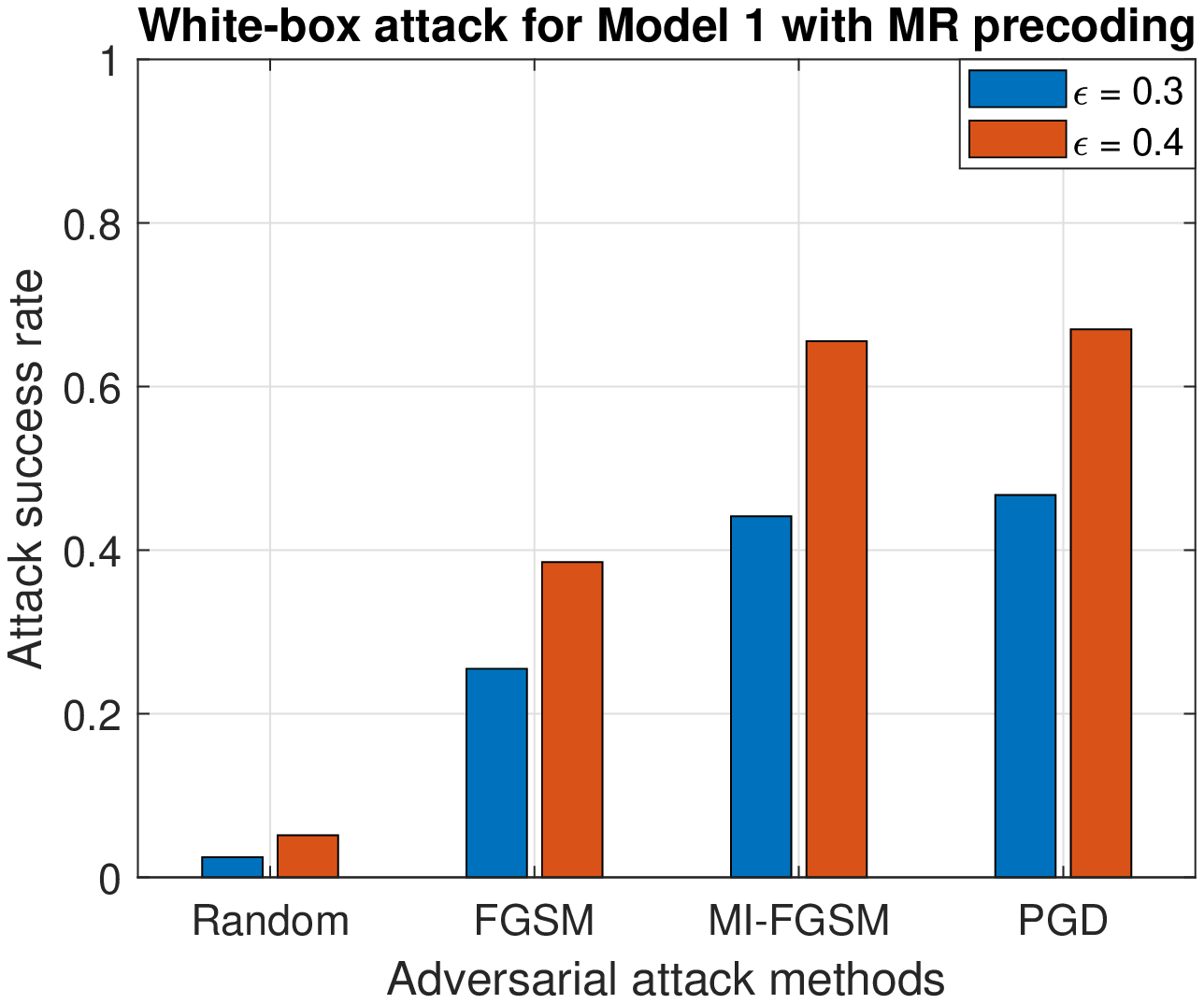}}
  \centerline{(a)}\medskip
\end{minipage}
\begin{minipage}[b]{.493\linewidth}
  \centering
  \centerline{\includegraphics[width=1.13\linewidth]{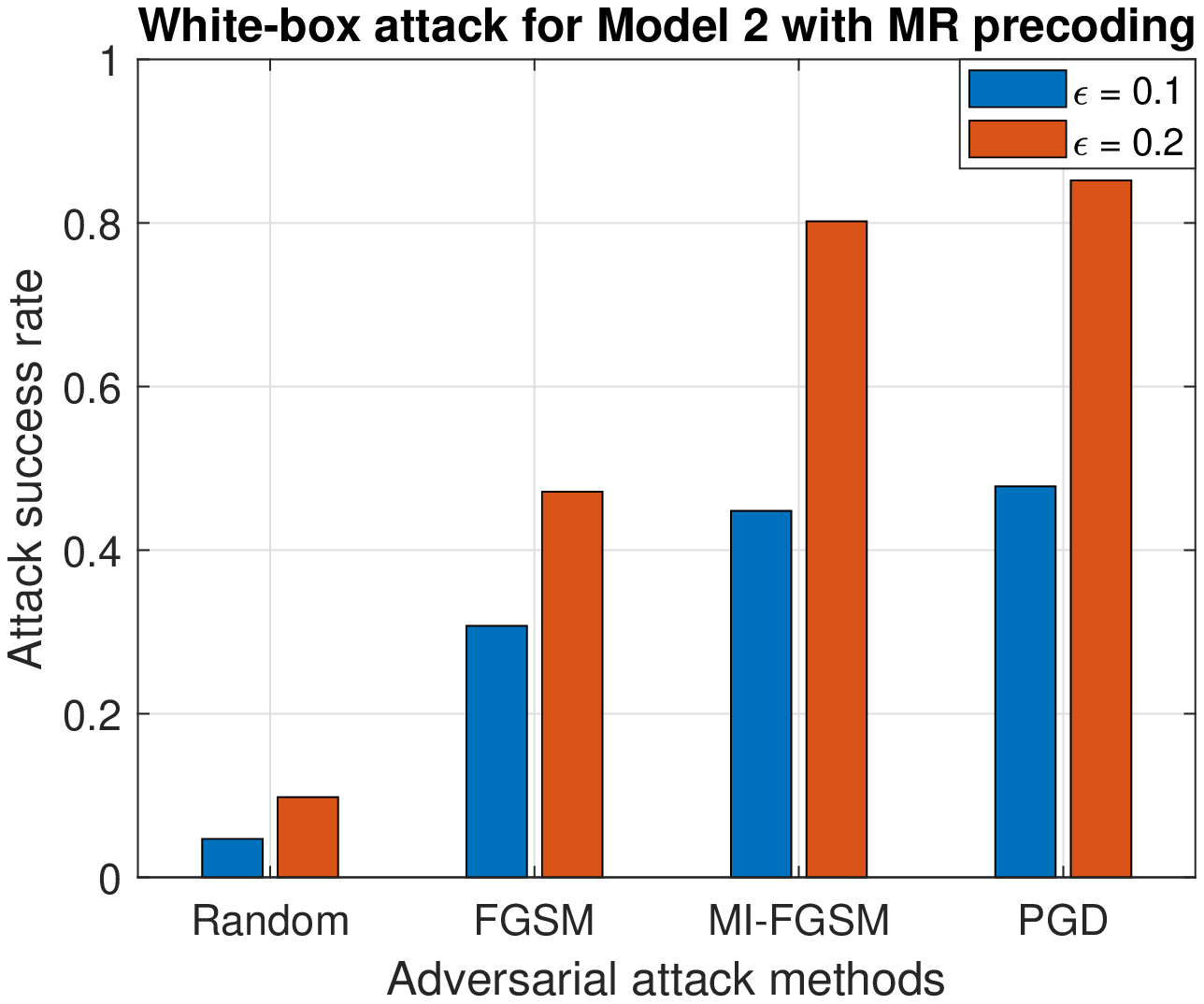}}
  \centerline{(b)}\medskip
\end{minipage}
\hfill
\begin{minipage}[b]{0.493\linewidth}
  \centering
  \centerline{\includegraphics[width=1.1\linewidth]{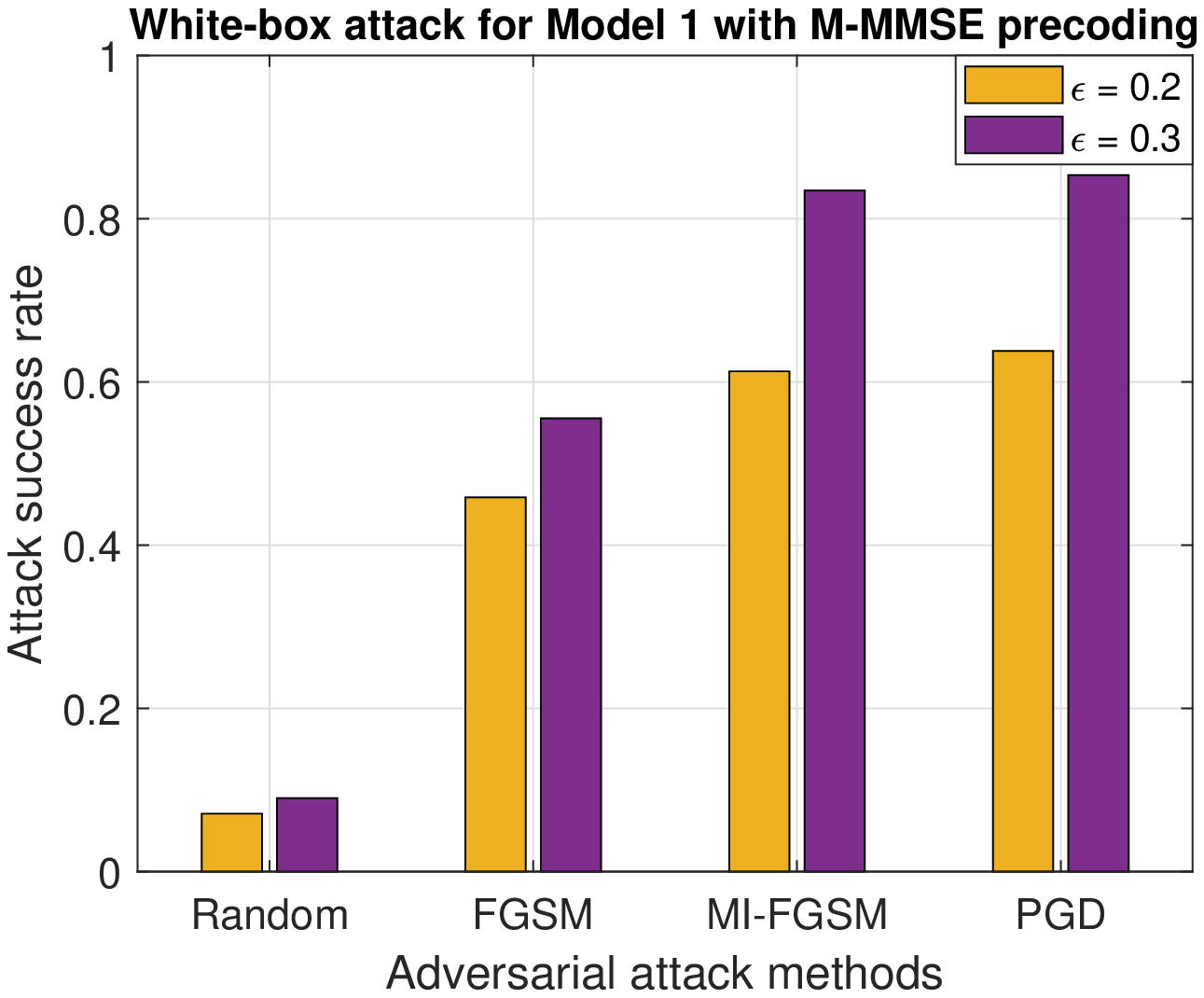}}
  \centerline{(c)}\medskip
\end{minipage}
\begin{minipage}[b]{0.493\linewidth}
  \centering
  \centerline{\includegraphics[width=1.1\linewidth]{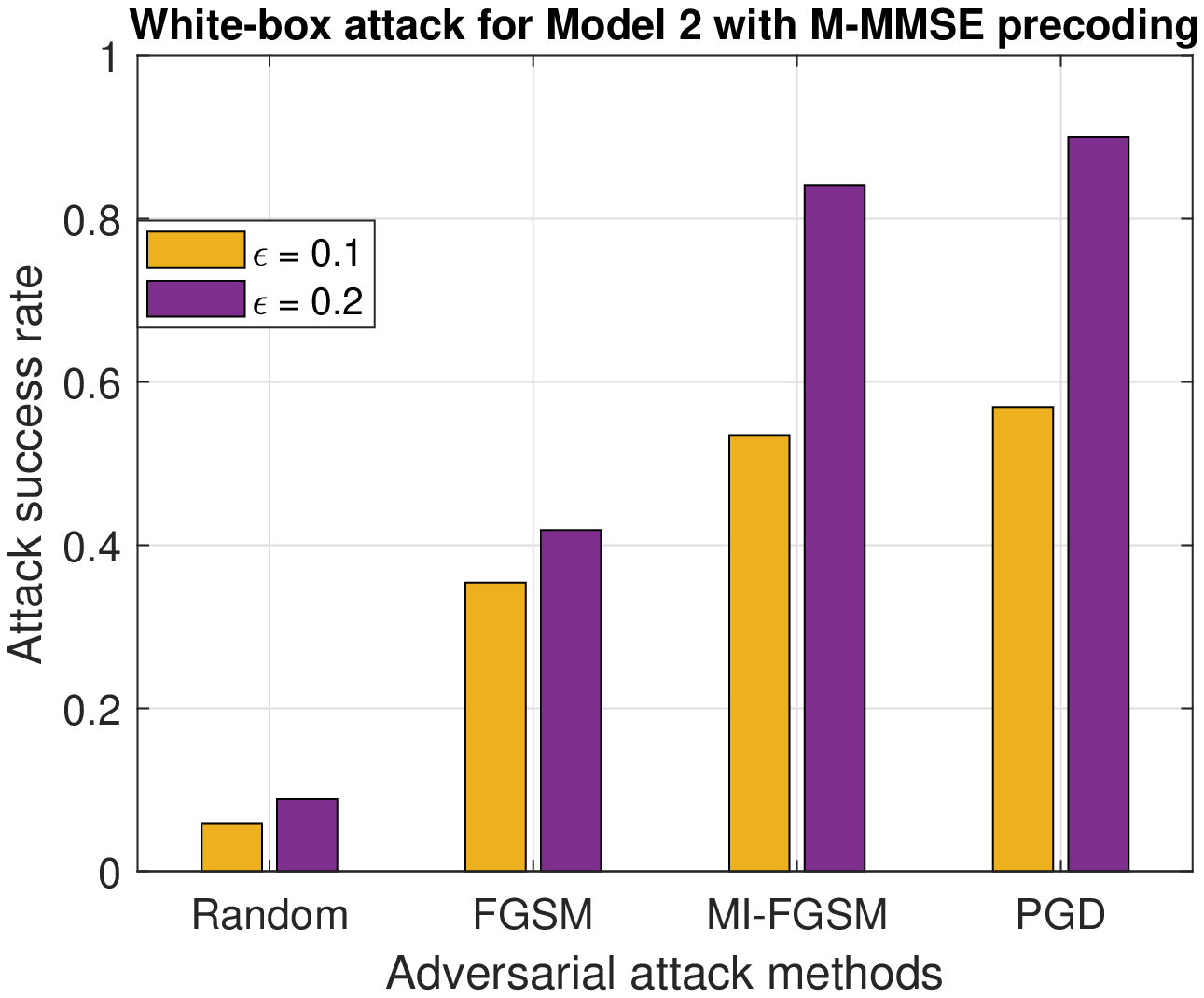}}
  \centerline{(d)}\medskip
\end{minipage}
\caption{Success rate of different adversarial attack methods for white-box attacks (a)--(d).}
\label{fig:res1}
\end{figure}

Figs. \ref{fig:res1} (a) and (b) show the cell-based comparison of success rate of different adversarial attacks in fooling the DNN `Model 1' and `Model 2', respectively, that are trained with MR precoding scheme. The input samples to the DNN models that are considered in the experiments are random perturbation ${\bf{x}}_\text{rnd}$ generated with probability $\Pr(\epsilon) = \Pr(-\epsilon) = 0.5$ and adversarial perturbation ${\bf{x}}_\text{adv}$. The experiments are conducted with very small perturbation magnitudes of $\epsilon = 0.3$ and $\epsilon = 0.4$  for `Model 1' with corresponding distance perturbations of UEs positions by $d_\epsilon = 42.4$cm and $d_\epsilon = 56.5$cm in the gradient direction to increase the loss; $\epsilon = 0.1$ and $\epsilon = 0.2$  for `Model 2' with corresponding distance perturbations of $d_\epsilon = 14.1$cm and $d_\epsilon = 28.2$cm.  
It can be observed from the  
figure that with the increase in $\epsilon$, the number of samples providing infeasible solution also increases.  
From Fig. \ref{fig:res1} (a) for DNN `Model 1' under MR precoding with $\epsilon = 0.3$, the percentage of infeasible solution across all cells provided by random perturbation is just 2\%, while PGD attack achieves 46\%.  
For DNN `Model 1' under MR precoding with $\epsilon = 0.4$, 
the minimum and maximum success rate of fooling the NN is
5\% and 67\% for random perturbation and PGD attack, respectively. From Fig. \ref{fig:res1} (b), for DNN `Model 2' under MR precoding with $\epsilon = 0.1$, the success rate of fooling the NN is lowest for random perturbation with 4\% and highest for PGD attack with 47\%; for  $\epsilon = 0.2$, the fooling rate reaches 85\% for PGD attack and 9\% for random perturbation.
Figs. \ref{fig:res1} (c) and (d) present the infeasible power solution obtained due to different methods of adversarial perturbations on DNN `Model 1' and `Model 2', that are trained with M-MMSE precoding scheme. From Fig. \ref{fig:res1} (c), for DNN `Model 1' under M-MMSE precoding with $\epsilon = 0.2$, the PGD attack can achieve 63\% and only 7\% for random white noise as success rates of fooling the network; for $\epsilon = 0.3$, the success rate of PGD attack achieves 85\%. 
From Fig. \ref{fig:res1} (d), for DNN `Model 2' under M-MMSE precoding, it can be seen that even for a small value of $\epsilon = 0.2$, for a majority of the input samples, the NN predicts infeasible output; the success rate of causing erroneous output in this case is 90\% for PGD attack and only 8\% for random perturbation. 

The DNN `Model 2' is a complex network that was proposed in \cite{Sanguinetti_dl_2018} in order to improve the learning abilities. Though `Model 2' is computationally complex NN consuming more time during the training phase, it is also more vulnerable for the adversarial attacks as it is evident from the the Figs. \ref{fig:res1} (b) and (d). This claim is consistent with the iterative-based gradient update attacks such as MI-FGSM and PGD, whereas this is not always true for the FGSM attack at least for the $\epsilon$ values that we have considered in our experiment. As a final note, we conclude that in comparison to FGSM and MI-FGSM attacks, the PGD method is a more powerful adversarial attack as it  consistently shows high success rate in fooling the NNs for all values of $\epsilon$ which can be observed from the Figs. \ref{fig:res1} (a)--(d). 
\subsection{Black-box attacks}
In this subsection, for the considered regression-based models, we analyze the success rate of different adversarial attacks to fool the DNNs using transferability. In other words, we analyze how well the adversarial attacks are transferable for our setting. To this end, we assume that the adversary has complete knowledge of the  `Model 1', i.e., the adversary has access to the structure of the NN. Thus, the adversary can train this substitute `Model 1' and generate the adversarial examples using FGSM, MI-FGSM, and PGD attacks. To analyze transferability, we assume that the adversary does not have access to the DNN `Model 2' and hence is the victim model; this model is investigated for the vulnerability against the adversarial examples crafted using substitute `Model 1'. We have generated the dataset individually   
with MR and M-MMSE precoding schemes that produces only feasible output solution on the clean samples for the DNN `Model 1' and `Model 2'. 

\begin{figure}[!htp]
\begin{minipage}[b]{0.493\linewidth}
  \centering
  \centerline{\includegraphics[width=1.13\linewidth]{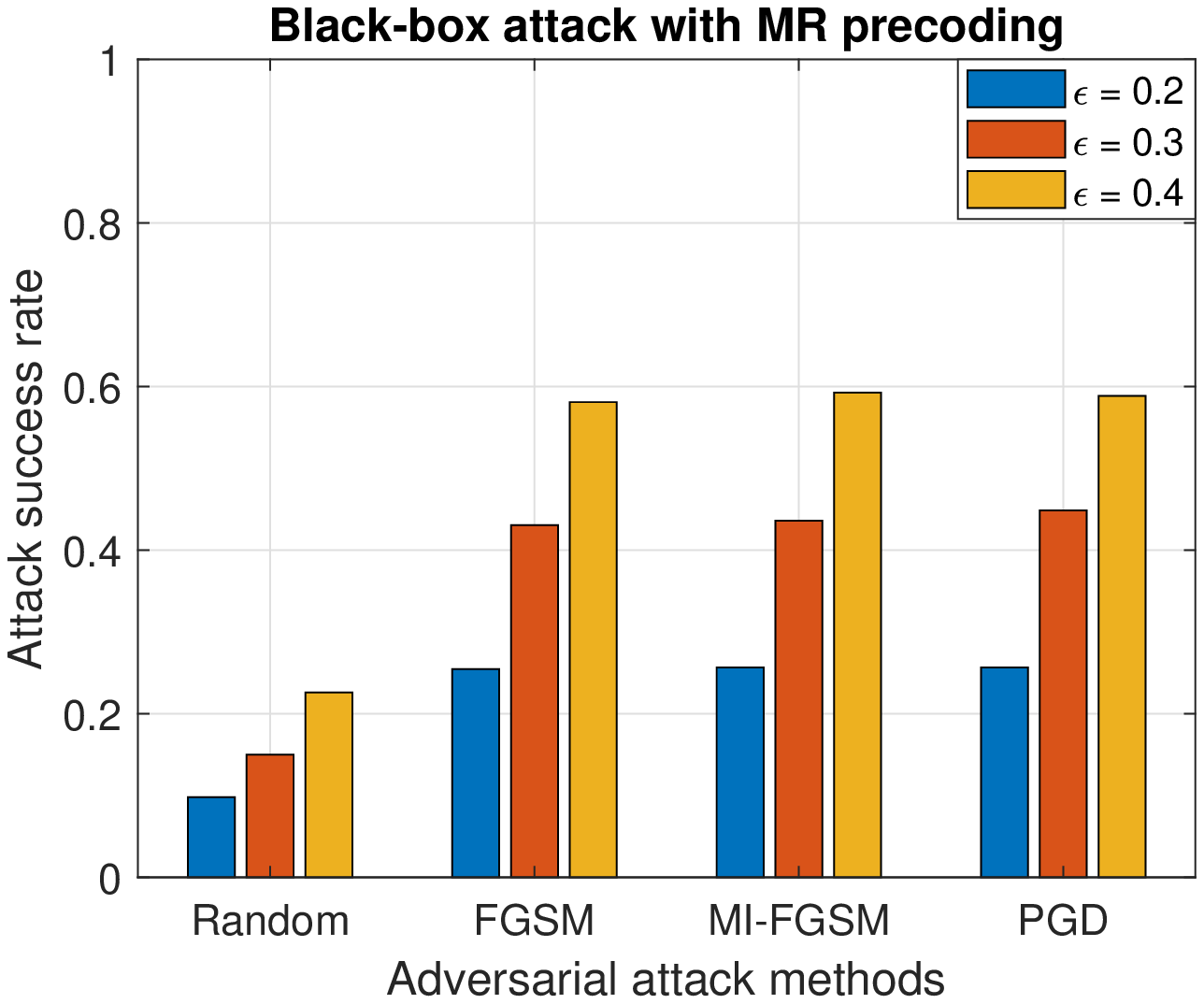}}
  \centerline{(a)}\medskip
\end{minipage}
\begin{minipage}[b]{0.493\linewidth}
  \centering
  \centerline{\includegraphics[width=1.13\linewidth]{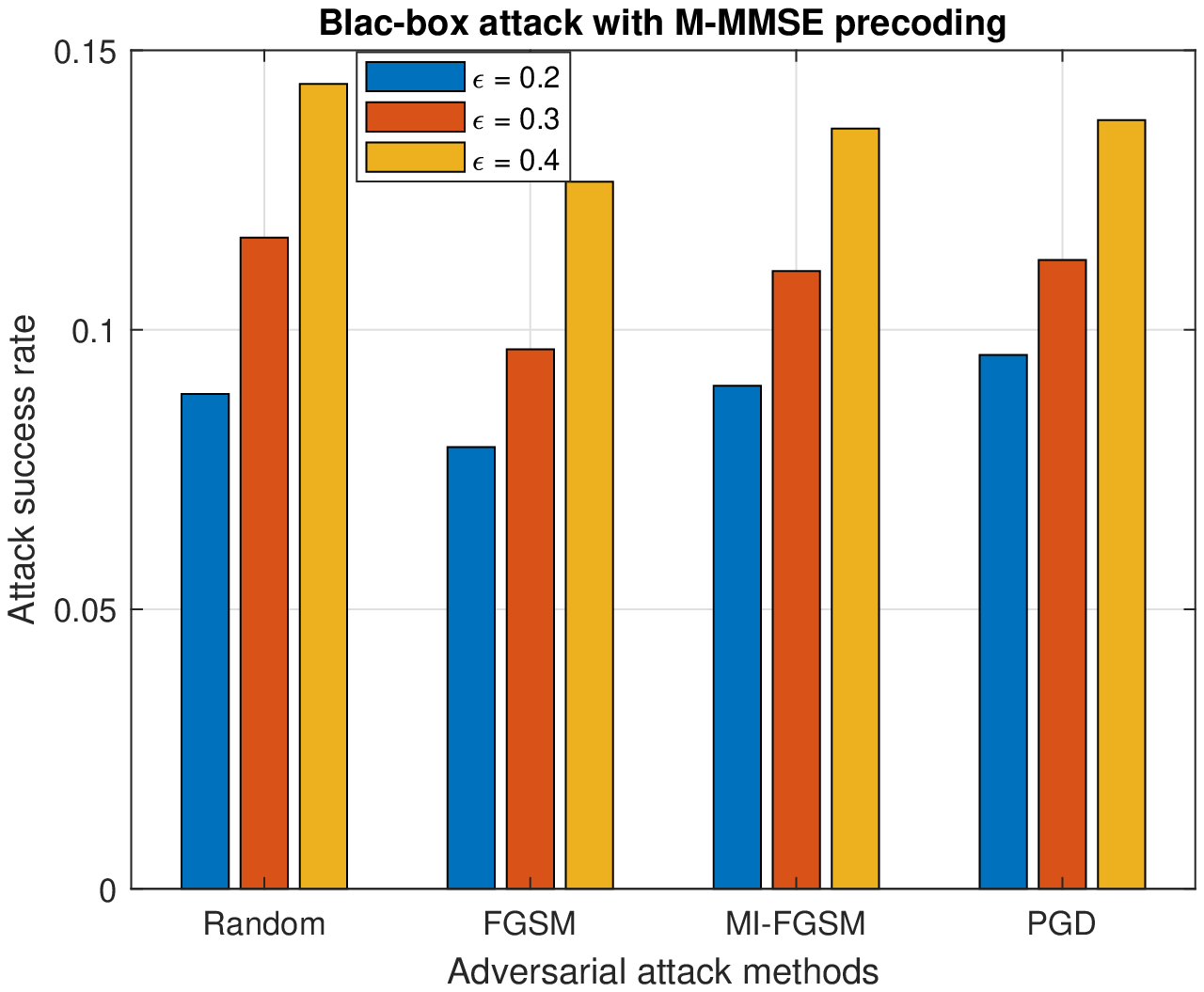}}
  \centerline{(b)}\medskip
\end{minipage}
\caption{Success rate of different adversarial attack methods for  black-box attacks (a)--(b) with surrogate `Model 1' and victim `Model 2'.}
\label{fig:res2}
\end{figure}

Figs. \ref{fig:res2} (a) and (b) present the infeasible solution that is obtained 
with MR and M-MMSE precoding schemes, respectively, depicting the black-box attacks using different adversarial methods. 
From Fig. \ref{fig:res2} (a), we have the success rate of fooling the NN using transferability for $\epsilon = 0.2$ as 25\% for all the attacks and 9\% for random perturbation; for $\epsilon = 0.3$, it can be achieved a maximum of 44\% using  PGD attack; for $\epsilon = 0.4$, 59\% can be achieved using MI-FGSM and 22\% for random perturbation. The percentage suggests that for the network trained under MR precoding, the success rate of transferability with small values of $\epsilon$ is equally good for both one-step and iterative gradient update attacks. However, the success rate of fooling the network using black-box attacks for random perturbation is also high as compared to the case of white-box attacks. 
From Fig. \ref{fig:res2} (b), we observe that the success rate of fooling the NN with M-MMSE precoding scheme for $\epsilon = 0.2$ is 9\% for both PGD attack and random perturbation; 
for $\epsilon = 0.3$, there is not much improvement in fooling the network, it is only  11\% for both PGD and random perturbation; 
for $\epsilon = 0.4$, the fooling rate due to random perturbation of 14\% is marginally higher than the other attack methods with PGD attack having 13\%. 
The results from this figure suggest that, the NN trained under M-MMSE precoding has poor transferability. In fact, the percentage of infeasible solutions obtained for random perturbations is more than the adversarial perturbations. Figs. \ref{fig:res2} (a) and (b) suggest that the adversarial success rate for the victim `Model 2' with MR precoding scheme is higher than the M-MMSE precoding scheme.  
\section{Conclusion}
In this paper, by considering the dataset of max-product SINR optimal power allocation strategy in a multi-cell maMIMO network with MR and M-MMSE precoding schemes, we have proposed different ways to craft the adversarial examples using the gradient-based methods for the regression-based DNN models. In white-box attack, we found that the PGD adversary attack has the highest success rate in fooling the NN as compared to that of FGSM and MI-FGSM attacks. The number of gradient update iterations to increase the loss function in order to craft adversarial examples in MI-FGSM is less than the PGD, thus, MI-FGSM is efficient in terms of computational complexity with a success rate higher than the FGSM and slightly lesser than the PGD attack method. In black-box attacks, we found that the NN `Model 2' trained under MR precoding scheme is more vulnerable than the NN trained under M-MMSE precoding scheme. 

\end{document}